# Nonlinear transport of Wigner solid phase surrounding the two-flux composite fermion liquid

Yu-jiang Dong, Xinghao Wang, Jianmin Zheng, Weiliang Qiao, and Rui-Rui Du*
*International Center for Quantum Materials, Peking University, Beijing 100871, China*

Loren N. Pfeiffer, Kenneth W. West, and Kirk W. Baldwin
*Department of Electrical Engineering, Princeton University, Princeton, New Jersey 08544, USA*

We have investigated the low temperature ($T$) transport properties of fractional quantum Hall (FQH) states in a high-mobility two-dimensional hole gas. According to the composite fermion (CF) model, FQH states stemming from a half-filled Landau level, specifically at fillinsg factors $\nu = p/(2p+1)$ ($p = \pm 1, \pm 2, \pm 3, \ldots$), can be associated with two-flux-attached CFs at the corresponding Lambda filling factor $p$. The zero-resistance minima and Hall plateaus of these states exhibit unusual temperature dependencies, characterized by rapid increases in width below a threshold temperature around 100 mK. Differential conductivity measurements from Corbino samples reveal that the regimes surrounding the CF liquid display clear nonlinear transport characteristics. This nonlinearity implies that each CF liquid is surrounded by CF solid phase composed of dilute CF excitations. Quantitatively, the applied electric field $E$ influences the motion of CF solid in a way analogous to $T$, which is dubbed the "$E-T$ duality." Our analysis indicates that this $E-T$ duality is consistent with the Berezinskii-Kosterlitz-Thouless theory in two-dimensional phase transitions.

*Introduction.* Under sufficiently low temperature ($T$) and strong perpendicular magnetic field ($B$), clean two dimensional (2D) electron/hole systems exhibit rich quantum phases as their ground states, including the fractional quantum Hall (FQH) states at odd-denominator filling factors, as well as those at even-denominator filling factors such as $\nu = 5/2$ and 7/2 [1–3]. Additionally, symmetry-broken phases like Wigner solids (WSs) [4–9], nematic stripe phases, and bubble phases [10–17] can also emerge. It is by now well established that the formation of these ground states is driven by Coulomb interactions, with the interaction properties prescribed by the pseudopotentials at different Landau levels (LLs). Remarkably, the correlation of electrons in the LLs can be accurately described by the composite fermion (CF) theory [18,19].

The competition between FQH states and WSs has attracted extensive attention over the past few decades. An ideal 2D system is expected to condense into a WS at sufficiently low LL filling factor $\nu = nh/eB$, where $n$ represents the carrier sheet density [20,21]. Additionally, a parameter $\kappa$ is conventionally defined to describe LL mixing, which also plays a crucial role in determining the pseudopotentials at finite $B$. Specifically, $\kappa = E_c/\hbar\omega_c = (e^2/4\pi\epsilon l_B)/(\hbar eB/m^*)$ represents the ratio between the Coulomb interaction energy $E_C$ and the cyclotron energy $\hbar\omega_c$. ($l_B = \sqrt{\hbar/eB}$ is the magnetic length, $\epsilon$ is the dielectric constant, $m^*$ is the effective mass.) Theoretical calculations favor a WS ground state when $\kappa$ is sufficiently large [22–24].

In the case of 2D electron gas (2DEG) in GaAs, with an effective mass $m_e^* \sim 0.067\, m_e$, $\kappa$ is relatively small, leading to the expectation that a WS will dominate only at filling factors $\nu \ll 1/5$. Transport experiments have observed insulating phases (IPs) near the 1/5 FQH state [5,6,25,26]. Those phases are believed to be WSs pinned by disorders, as confirmed by, e.g., nonlinear electric response [6,7,25] and microwave resonances [7,8,27,28]. For the 2D hole gas (2DHG) in GaAs, which is the subject of this work, the effective mass $m_p^* \sim 0.4 m_e$ is significantly larger, resulting in a larger $\kappa$ that promotes the formation of a WS at filling factors $\nu < 1/3$ [23,24,29,30].

A WS can be formed not only by electrons/holes, but also by composite fermions (CFs) [31–34]. According to the CF theory, the Jain-sequence FQH states [$\nu = p/(2mp+1)$] can be mapped to integer quantum Hall states (IQHSs) of CFs at the Lambda level ($\Lambda$) filling factor $p$, by attaching $2m$ flux quanta to each electron/hole (termed CF-$2m$) [18]. In this Letter, we focus exclusively on the case of $m = 1$. For instance, the $\nu = 1$ state maps to $p = -1$, and $\nu = 1/3$ maps to $p = 1$. The WS of CFs (CFWS) existing at very low filling factors is usually called a "type-I CFWS," and is believed to be more stable than electron WSs between 1/5 and 2/9, as supported by numerical calculations [24]. On the flanks of each $\Lambda$ level, a dilute gas of interacting CFs forms a WS known as a "type-II CFWS" which coexists with the FQH liquid constituted by the majority of CFs [35]. Thermal activation [36], surface acoustic wave [37], and microwave resonance [38] experiments have confirmed the existence of the type-II CFWS around $p = -1$ and $p = 1$. Additionally, theoretical prediction suggests that the 2/5 ($p = 2$) and 3/7 ($p = 3$) states also contain narrower ranges of CFWSs [35].

---

*Contact author: rrd@pku.edu.cn

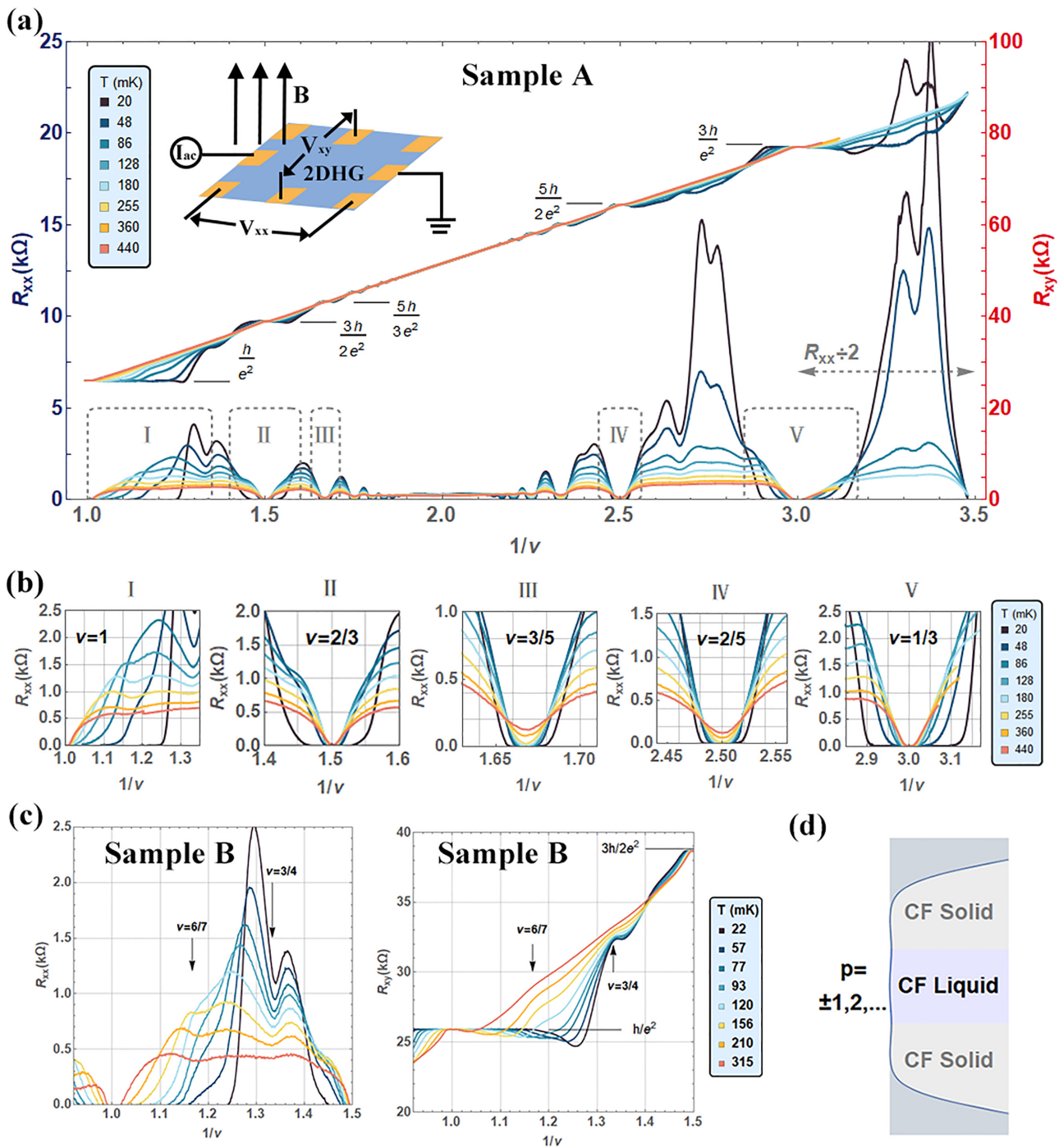

FIG. 1. (a) $R_{xx}$ and $R_{xy}$ of vdP sample A under different temperatures. The horizontal axis is rescaled by the reciprocal of the filling factor $1/\nu$ to highlight each FQH state. The inset shows the measurement configuration of our sample. (b) The $R_{xx}$ of the five major FQH states are magnified to illustrate the transition from U shape to V shape as temperature increases [respectively corresponding to the dashed gray boxes in (a)]. (c) $R_{xx}$ and $R_{xy}$ of vdP sample B, which has higher density and mobility. (d) Schematic of the differential conductance in type-II CFWS, where the width is broadened at low temperature due to CF solid phases at the flanks of the CF liquid.

Here, we report low-temperature transport experimental results on the flanks of the FQH series $\nu = p/(2p + 1)$, $p = \pm 1, \pm 2, \ldots$ in a high-mobility 2DHG, and analyze the data based on the CFWS picture. In differential conductivity measurements from Corbino samples, we observed that the regimes surrounding the CF liquid exhibit clear nonlinear transport characteristics, indicative of threshold behavior. We infer that CF liquid is surrounded by CF solid. Interestingly, the applied electric field $E$ influences the melting transition of the solids in a way analogous to the temperature $T$, leading to the conjecture of "$E-T$ duality" [34]. Our analysis indicates that this $E-T$ duality is consistent with the Berezinskii-Kosterlitz-Thouless (BKT) theory in two-dimensional phase transitions [39].

*Magnetotransport measurement.* Our experiments were carried out using carbon-doped high-mobility GaAs/Al$_{0.33}$Ga$_{0.67}$As narrow (20-nm-wide) quantum wells (QWs). Transport measurement was performed in a dilution refrigerator (Oxford KelvinTLM-18T) with a base temperature of 20 mK and a superconductor magnet. Samples A and C are respectively a van de Pauw (vdP) and a Corbino device, both fabricated from a QW wafer with a hole density of $1.21 \times 10^{11}$ cm$^{-2}$ and mobility of approximately $1.4 \times 10^6$ cm$^2$/Vs. Sample B is a vdP device fabricated from a QW wafer with a hole density of $1.47 \times 10^{11}$ cm$^{-2}$ and mobility of approximately $2.18 \times 10^6$ cm$^2$/Vs.

An overview of the magnetotransport properties of vdP sample A is shown in Fig. 1(a). At high magnetic filling factors less than 1/3, the $R_{xy}$ trace is slightly mixed with $R_{xx}$ due to vdP configuration. Figure 1(b) shows a detailed evolution of $R_{xx}$ for the five major CF states at different temperatures. Notably, both the widths of the zero resistance states and the corresponding $R_{xy}$ plateaus broaden significantly as the temperature decreases from 440 mK (red line) to 20 mK (black line), accompanied by a transition of the $R_{xx}$ minima from "V shape" to "U shape."

This V-shape to U-shape transition can be attributed to the solidification of the WS [40], as the reduction in longitudinal conductance $\sigma_{xx}$ ($\propto R_{xx}$) signifies the emergence of insulating states. According to the depiction of type-II CFWS, here CF liquid phases occupy the centers of the CF filling factors, while at the flanks the remaining dilute carriers form CF solid phases [35] [Fig. 1(d)]. Since the CF solids at the flanks do not contribute to conductance, $R_{xy}$ will remain quantized while $R_{xx}$ will stay at zero, resulting in a U shape at low temperature. A similar understanding of the width at the $\nu = 1$ QH state in a very-high mobility 2DEG has been presented in Ref. [36], albeit there the phases are described by electrons rather than CFs. In Ref. [37], the existence of a pinned WS near $\nu = 1$ and $\nu = 1/3$ in a 2DHG is also substantiated. Our result is a step further to say that in a high-mobility, low-density 2DHG, a CFWSs may exist around all the five CF-2 states shown in Fig. 1(b).

It is noteworthy that the right wing of $1/\nu = 1$ exhibits distinct characteristics compared to other states. Increasing the temperature beyond 128 mK, the resistance at approximately $1/\nu = 1.17$ and 1.33 evolves into a shallow $R_{xx}$ minimum. This observation is consistent with those presented in Ref. [41]. To better resolve the details, we measured an additional vdP sample (sample B) with higher density and mobility, plotting the data from $1/\nu = 0.8$ to $1/\nu = 1.5$ [Fig. 1(c)]. The $1/\nu > 1$ region qualitatively matches the results presented in Fig. 1(a), and both the $R_{xy}$ plateau and the $R_{xx}$ minimum indeed exhibit asymmetry with respect to $\nu = 1$.

We attribute this asymmetry at filling factor $1/\nu > 1.14$ to the emergent states, such as a developing FQH state at 3/4 [41], and a potential FQH state near 6/7 [42]. To concentrate on the transport properties of the CFWS surrounding the $p = -1$ CF liquid, we will analyze the data within a range $1/\nu = 0.92$ to 1.14 in the subsequent discussion. It should be noted that slight asymmetry can also be observed for both the filling factors of 1/3 and 2/3. For the current samples, evaluating the influence of emergent minima on the flanks of these fractions remains less straightforward.

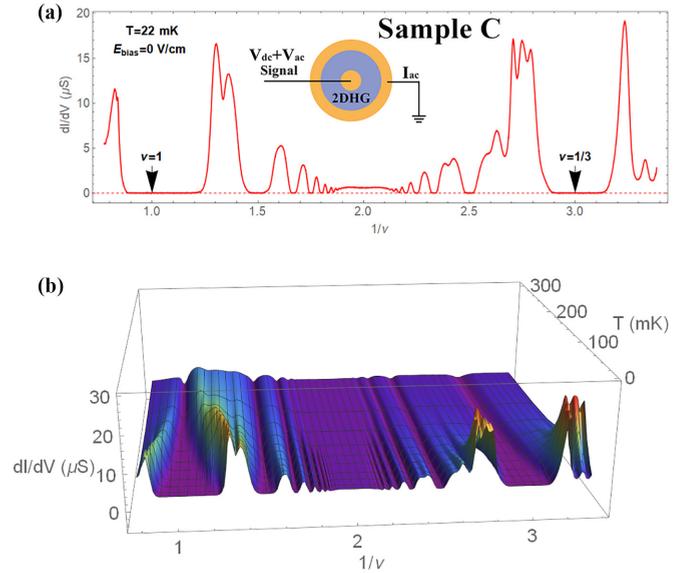

FIG. 2. (a) Differential conductance at zero bias electric field of sample C at a base temperature of 20 mK. $\nu = 1$ and $\nu = 1/3$ are marked by arrows. The inset provides a schematic of Corbino measurement circuit. (b) The 3D plot of the longitudinal conductivity $dI/dV$ at zero bias electric field ranging $1/\nu = 0.77$ to 3.38, with $T$ from 20 to 300 mK.

*Nonlinear differential conductance.* A hallmark transport characteristic of a pinned WS is its nonlinear behavior, which has been interpreted as resulting from the sliding motion of pinned solids [6,7,16,34,43,44]. In this model, multiple solid domains are pinned by the disorder potential and remain stationary under a low voltage bias. However, when a sufficiently strong external electric field is applied, these domains can surmount the pinning energy barrier and contribute to the finite conductance [45]. This model is widely applied in other physical systems, including type-II superconductors and magnetic skyrmions [46].

Therefore, a Corbino sample C was fabricated to investigate its nonlinear properties. It was fabricated from the same wafer as sample A, featuring outer and inner contact diameters of approximately 3500 and 700 μm respectively. Figure 2(a) shows the differential conductance $dI/dV$ of sample C versus $1/\nu$ at zero bias electric field. The mean differential conductance around the main CF-2 state is $\langle dI/dV \rangle < 6$ nS, indicating a zero-conductance state (ZCS).

Figure 2(b) presents a 3D plot illustrating the relationship between a $dI/dV$ trace and different temperatures ($T$). ZCS valleys are depicted in purple, while conductance peaks representing the melted solids are colored red. At integer fillings of Λ levels, CF liquid is consistently observed across the entire temperature range $20 - 300$ mK, while the flanks shrink rapidly. These key characteristics are in agreement with the vdP measurement results, ensuring consistency across samples and configurations.

In the following nonlinear measurement, we apply an external dc electric field ($E_{bias}$) using a combined dc + ac voltage source. Although $E_{bias}$ is a logarithmic function of the radius $r$, it can be approximated by the average electric field $\langle E \rangle$, which is estimated as the ratio of voltage difference ($\Delta V_{DC}$) and radius difference of the two contacts $\Delta R$. Specifically, $\langle E \rangle = \int_{R_{in}}^{R_{out}} E(r)dr/(R_{out} - R_{in}) = \Delta V_{DC}/\Delta R$. The measured AC current $dI$, obtained via lock-in technique, is divided by the input AC voltage $dV$ to determine the differential conductance $dI/dV$.

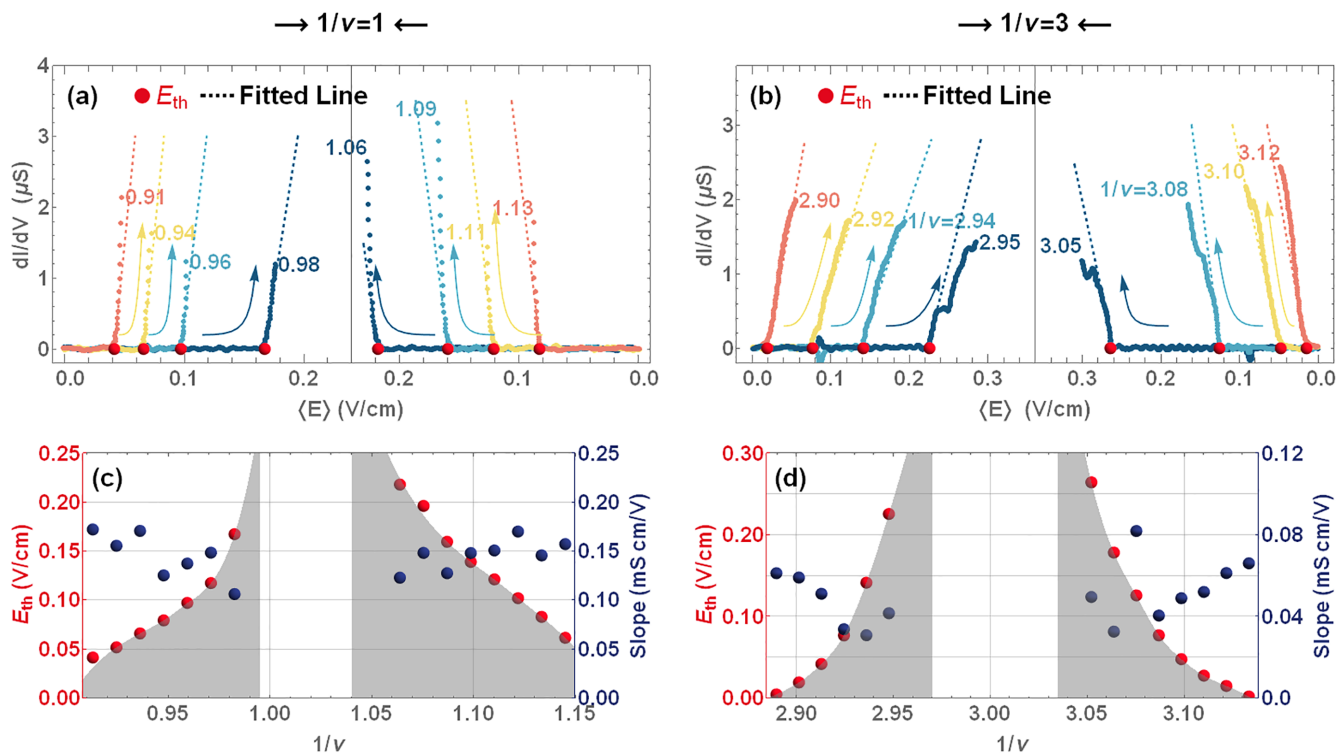

FIG. 3. (a), (b) Nonlinear differential conductance around $\nu = 1$ and $\nu = 1/3$ vs external electric field. Cooler colors indicate the proximity to the center of CF states and stronger WS, warmer color vice versa. The reciprocals of filling factors ($1/\nu$) are labeled beside each line. The slopes of the rising $dI/dV$ trace are fitted by dashed lines. The threshold electric fields ($E_{th}$), marked by red points, are determined by $dI/dV < 0.1$ μS. (c), (d) The threshold fields (red dots) and slopes (blue dots) vs $1/\nu$ around $\nu = 1$ and $\nu = 1/3$. Solid phases measured in our experiment are colored in gray.

Regions from $1/\nu = 0.87$ to 3.17 are selected to apply electric field to observe the changes in differential conductance $dI/dV$. At the two wings of CF-2 states, $dI/dV$ exhibits a sharp transition from zero to finite values as $\langle E \rangle$ increases, indicating a nonlinear response [45–47] [Figs. 3(a) and 3(b)]. Among all the CF-2 states, IPs near $p = \pm 1$ (i.e., $\nu = 1$ and $1/3$) demonstrate the most pronounced nonlinear response. Therefore, the following analysis will concentrate on the flanks of these two states.

We plot the data from states around $\nu = 1$ in Fig. 3(a). As the filling factor approaches 1, the solid states become stronger with a larger shear modulus, which we represent with colder colors. The threshold electric fields ($E_{th}$), indicated by red points, are determined by $dI/dV = 0.1$ μS. These points define the boundaries of the CF solid phase. Additionally, we fit the increasing slopes of $dI/dV$ beyond $E_{th}$, as shown by the dashed lines in Fig. 3(a). Since the slopes vary gradually with the bias voltage, we only use the data points near threshold positions for the fitting.

To more clearly illustrate the variation trends of the melting-transition parameters with filling factors, we specially plot $E_{th}$ (red dots) and the slopes (blue dots) in Fig. 3(c). It is not surprising that CF solid exhibits larger $E_{th}$ as the filling factor approaches CF liquid. This behavior can be attributed to the lower excess carrier density in the solid phase, which is proportional to $\Delta \nu = |\nu - 1|$. At the exact filling factor 1, $dI/dV$ remains zero throughout, indicating a linear response characteristic of Fermi liquid. This observation aligns with general type-II CFWS theory. Consequently, the CF solid phases can be identified based on the nonlinear response regime, where the regions correspond to $1/\nu$ values from 0.92 to 0.995 and from 1.04 to 1.14.

The states around $\nu = 1/3$ are also analyzed using the same methodology as those around $\nu = 1$, as illustrated in Figs. 3(b) and 3(d). For the CFWS here, the phase boundary is defined by $1/\nu$ ranging 2.89–2.97 and 3.03–3.13. These findings are consistent with the microwave resonance experiment reported in Ref. [38]. As illustrated in Fig. 3, the slopes of the rising conductivity exhibit a slight downward trend towards the center for the data around $\nu = 1$ (although accompanied by some degree of fluctuations), and more clear downward variations toward $\nu = 1/3$. Beyond the threshold, the nonlinear transport becomes increasingly complex, i.e., it may involve multiple phases such as solid-liquid mixtures. Theoretically, various mechanisms may be required to explain these phenomena, including depinning transition [46], melting transition [40], and possible contributions from percolations via variable range hopping [48]. Systematic experiments are certainly warranted to further investigate nonlinear transport beyond the threshold. In this study, we will focus on the primary findings, specifically the threshold behavior in CFWS transport.

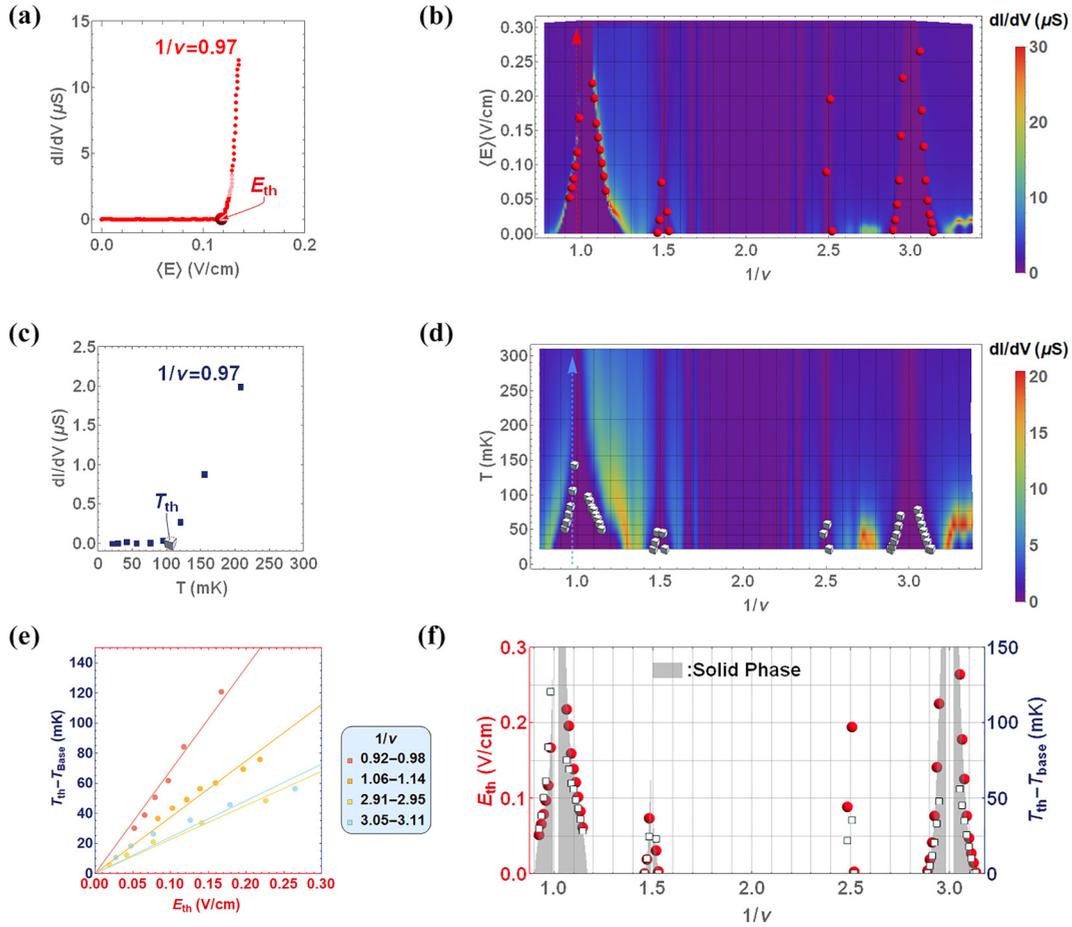

FIG. 4. (a) As an example, the electric field nonlinear response is shown at $1/\nu = 0.97$, extracted from the cutting line in (b). Threshold electric field $E_{\text{th}}$ is indicated by a red dot. (b) 2D differential conductance map as a function of $\langle E \rangle$ and $1/\nu$; threshold electric fields $E_{\text{th}}$ are indicated by red dots. (c) Temperature nonlinear response shown at $1/\nu = 0.97$, extracted from the cutting line in (d). Threshold temperature $T_{\text{th}}$ is indicated by a gray cube. (d) 2D differential conductance map as a function of $T$ and $1/\nu$; threshold temperatures $T_{\text{th}}$ are indicated by gray cubes. (e) Threshold temperature $T_{\text{th}}$ vs threshold electric field $E_{\text{th}}$ from different filling factors. (f) $E_{\text{th}}$ (dark red dot) coincides with $T_{\text{th}}$ (gray cubes) at main CF-2 states.

*E-T duality.* We then map the $dI/dV - \langle E \rangle$ traces under various $1/\nu$ into a single diagram to illustrate the distribution of solid phases [Fig. 4(b)]. Purple regions denote ZCSs, where CF liquid and CF solid coexist, while red regions indicate conductance maxima, corresponding to the completion of the melting transition. Red dots mark the threshold fields $E_{\text{th}}$. Data along a cutting line at $1/\nu = 0.97$ are presented in Fig. 4(a).

We also map the $dI/dV - T$ traces under various $1/\nu$ [Fig. 4(d)], and mark the threshold temperature $T_{\text{th}}$ with gray cubes. The data along the cutting line at $1/\nu = 0.97$ is plotted in Fig. 4(c). This temperature trace exhibits similar nonlinear behavior to the $E_{\text{bias}}$ trace, and the threshold temperature $T_{\text{th}}$ is determined by $dI/dV = 0.1\ \mu$S using the same method as for the $E_{\text{th}}$.

If we superimpose the $E$ diagram and the $T$ diagram, a notable similarity in their outlines can be observed [Fig. 4(f)].

To further evaluate this phenomenon, we plot $T_{\text{th}}$ versus $E_{\text{th}}$ in Fig. 4(e). In this plot, data from different filling factors exhibit an approximate linear relationship between $E_{\text{th}}$ and $T_{\text{th}}$, albeit with different slopes. We refer to this similarity between the $T$ diagram and the $E$ diagram as the "$E-T$ duality," indicating that the electric field plays a role analogous to temperature in the phase transition. This phenomenon has been observed in WS phases near $\nu = 1/5$ in 2DHG [34].

We shall now discuss the appropriate physical picture for analyzing this $E-T$ duality. One potential approach is depinning theory, which focuses on field-induced nonlinear behavior at very low temperatures, characterized by $\sigma \propto (E - E_{\text{th}})^{\beta}$, where $\beta$ describes the elastic property of the sliding WS [46]. Temperature-induced nonlinear behavior, on the other hand, has received less attention. Some studies suggest that temperature acts as a thermal activation mechanism, leading to an empirical exponential relationship: $\sigma \propto \exp(-\frac{E_a}{k_B T})$, where $E_a$ is the activation energy [36]. Consequently, while it is feasible to analyze $E$ and $T$ independently, conventional depinning theory falls short in adequately addressing the duality of these two parameters. We propose that, in addition to the depinning transition, the melting of the WS under an electric field also contributes to the nonlinear transport phenomena observed in our experiments.

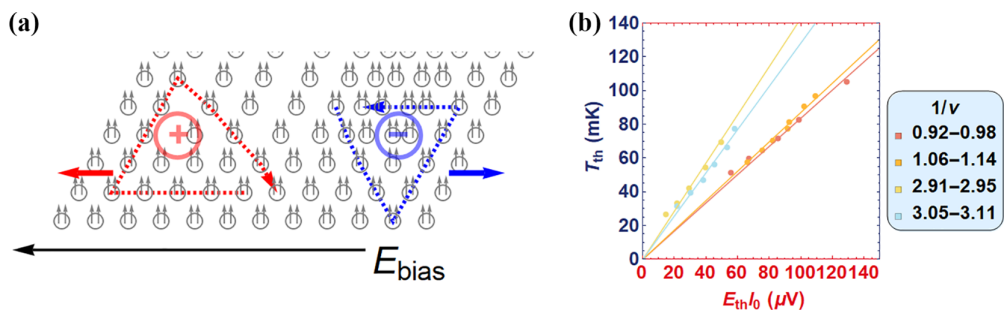

FIG. 5. (a) Sketch of topological defects in a CF-2 triangular lattice, and the role external electric field plays in our model. Positive (red) topological defect is labeled by a red (blue) circle. The paths defining our topological vortices are labeled by dashed arrows. (b) Linear fitted relationship between $T_{th}$ and $E_{th}l_0$.

Based on this analysis, we are motivated by the fact that the Berezinskii-Kosterlitz-Thouless (BKT) theory is also commonly applied to describe temperature-induced and field-induced nonlinear behavior in 2D systems, such as phase transitions in 2D superconductors [49]. It offers a different approach to elucidating the duality phenomenon. So, in the following section, we will discuss the *E-T* duality effect using BKT theory.

*Interpretation of data by the Berezinskii-Kosterlitz-Thouless (BKT) transition.* The 2D solid-liquid phase transition can be described according to BKT theory, which originates from the dislocation theory of melting 2D solids, and is extensively applied in the study of 2D phase transitions. According to BKT theory, while a conventional second-order phase transition is not possible in 2D systems, a topological phase transition driven by topological vortices can occur. In our system, these "topological vortices" manifest as defects that naturally arise during the melting of CF solids due to energy excitations. An example of a pair of topological defects is illustrated in Fig. 5(a), where an extra half row of CFs is inserted into (or removed from) the original lattice. Drawing an analogy to vortices in quantum liquids and quasiparticles in the FQH effect, it is appropriate to refer to these melting localized defects as "puddles."

According to BKT theory, isolated puddles are energetically unfavorable in the solid phase, due to their logarithmic increase in energy with the size of CF solid. Conversely, puddle-antipuddle pairs, or dipole configurations comprising oppositely charged puddles, possess finite energy and exhibit energetic stability. At sufficiently high temperature or under sufficiently large shear stress, puddles that initially belong to a single pair are split into two distinct, isolated puddles. This results in the transition of the quasi-long-range-ordered solid phase into a liquid phase lacking long-range order.

Let us first examine the effect of thermal excitations. According to BKT theory, the threshold temperature is proportional to shear modulus $C_T$, specifically $T_{th} = C_T/2\sqrt{3}k_B$, assuming that the CF solid forms a triangular lattice [35,39]. We compared the mean square separation of the puddles within a dipole pair $\langle r^2 \rangle$ with the mean square separation between pairs $\langle d^2 \rangle$. The ratio $\langle r^2/d^2 \rangle$ exhibits singularity at the threshold temperature $T_{th}$, indicating that the puddle dipoles become unstable at this temperature.

Based on the above model, we consider the effect of an external electric field $E$ by introducing an additional term $-e_d E r \cos\theta$ into the pairing energy, where $e_d$ is the effective charge of puddles and $\theta$ is the angle between $E$ and $r$. The external electric field contributes to the shear stress that splits a pair of puddles apart, leading to the transition from solid into liquid. Through numerical calculation (calculation details are shown in the Supplemental Material [50]), we find that the puddle pairs cannot remain a bounded state beyond a critical size $l_0$, which depends on the filling factor and base temperature. Since the shear modulus $C_T$ can be deduced from measurements of threshold temperatures, $l_0$ can be calculated at the base temperature accordingly. A more convincing relationship $E_{th}l_0 \propto T_{th}$ around $p = 1$ and $p = -1$ is shown in Fig. 5(b), with nearly identical slopes for filling factors near the same CF liquid state. This relationship elucidates the *E-T* duality effect mentioned earlier. The slope difference between $p = 1$ (slope $\sim$0.85 mK/$\mu$V) and $p = -1$ (slope $\sim$1.34 mK/$\mu$V) may arise from different defect charges of CF solid, presenting an interesting problem for further investigation.

The critical size $l_0$ deduced in this study can be interpreted as a characteristic length of domain size, approximately several micrometers. If the puddle distances exceed this value, the WS will undergo melting. This critical size is significantly larger than the domain size $L$ predicted by the conventional depinning model, which is determined by $L \sim k_B T_{th}/eE_{th}$, typically smaller than 1 $\mu$m. The latter can be measured through microwave experiments using the formula $L = (2\pi C_T/neBf_{pk})^{1/2}$, where $f_{pk}$ represents the resonance peak frequency [8].

To the best of our knowledge, no relevant microwave measurements have been reported for the CFWS in the context of 2DHG; it would be desirable to measure both the $L$ and $l_0$ in the same types of samples.

*Conclusion.* In summary, this work investigates the WS through nonlinear differential conductance measurements. We observe that type-II CFWS is prevalent around the most robust CF-2 Jain-sequence FQH states. Furthermore, in the context of the experiments where an electric field is applied in the 2DHG plane, we identify the duality between the threshold field $E_{th}$ and the threshold temperature $T_{th}$. By considering BKT theory, we propose a different perspective, that this duality can be explained by the approximate linear relationship between the shear modulus and the threshold electric field/temperature.

*Acknowledgments.* The work at PKU was funded by the National Key Research and Development Program of China (Grants No. 2019YFA0308400 and No. 2024YFA1409002), and by the Strategic Priority Research Program of Chinese Academy of Sciences (CN) (Grant No. XDB28000000). The work at Princeton was funded by the Gordon and Betty Moore Foundation through the EPiQS initiative Grant No. GBMF4420, and by the US National Science Foundation MRSEC Grant No. DMR-1420541.

*Data availability.* The data that support the findings of this article are not publicly available upon publication because it is not technically feasible and/or the cost of preparing, depositing, and hosting the data would be prohibitive within the terms of this research project. The data are available from the authors upon reasonable request.